\documentclass{elsart}

\usepackage{amssymb}
\usepackage{amsmath}
\journal{Physics Letters B}

\newcommand{\eps}{\varepsilon}
\newcommand{\ga}{\alpha}

\newcommand{\gd}{\delta}
\newcommand{\gl}{\lambda}
\newcommand{\gs}{\sigma}
\newcommand{\go}{\omega}

\renewcommand{\hat}{\widehat}

\newcommand{\rf}[1]{(\ref{#1})}

\begin{document}

\begin{frontmatter}

\begin{flushright}
hep-ph/0405104\\
ITFA-2004-19
\end{flushright}



\title{Tests of classical gravity description for microscopic black hole
production}


\author{Vyacheslav~S.~Rychkov}

\ead{rychkov@science.uva.nl}

\date{May 2004}

\address{Insituut voor
   Theoretische Fysica, Universiteit van Amsterdam\\
   Valckenierstraat 65, 1018XE Amsterdam, The Netherlands}

\begin{abstract}
The classical Einstein gravity description of black hole
production in transplanckian collisions in TeV-scale gravity is
tested for self-consistency. In addition to the ``curvature must
be small" test, which was shown to be violated in
[hep-ph/0401116], it is proposed to estimate quantum fluctuations
in the Aichelburg-Sexl shock waves corresponding to the colliding
particles. Using linearized quantum gravity, it is found that the
occupation numbers of gravitons with characteristic frequency are
too small to resolve the classical width of the shocks. This
raises further doubts in the classical gravity picture of black
hole creation and the geometric cross section estimate based on
it.
\end{abstract}

\begin{keyword}
Large extra dimensions \sep TeV-scale gravity \sep Transplanckian collisions
\PACS 04.70.-s \sep 04.50.+h \sep 11.10Kk
\end{keyword}
\end{frontmatter}

\section{Introduction}
\label{intro}

Microscopic classical black hole (BH) production in transplanckian
particle collisions is one of the most exciting possible
experimental signatures of large extra dimensions scenarios of
TeV-scale gravity (see \cite{Kanti} for a recent review). However,
thorough theoretical understanding of this process is still
lacking. Classical general relativity intuition tells us that
gravitational collapse should occur if the center-of-mass energy
of the colliding particles is deposited within a region of size
about the Schwarzschild radius corresponding to this energy
\cite{tHooft,BF,factory}. But is this intuition applicable in the
quantum world of elementary particles? In this Letter I will
present arguments suggesting that classical gravity description of
microscopic BH production may be inadequate.

The starting point of my discussion is the classical gravity
analysis of \cite{EG,YN}, which makes more precise the intuitive
picture described above.
These authors considered classical gravitational field of two fast
point particles in a grazing collision. They found that for impact
parameters of the order of the corresponding Schwarzschild radius,
the collision spacetime contains a closed trapped surface (CTS).
From this fact, BH formation follows by the singularity theorems
of classical general relativity and the cosmic censorship
conjecture.

My goal is to subject this analysis to validity checks, which have
to be passed by any classical field theory computation. The well-known
necessary conditions which have to be satisfied are: 1) field strengths
have to be small; 2) the number of field quanta has to be large.

The meaning of ``small'' and ``large'' in the preceding paragraph
depends on the situation and has to be decided on a case-by-case
basis. Familiar examples are provided by electromagnetism. The
first condition is violated when the electric field reaches the
critical value $\sim m_{\text{e}}^2/e$. The classical solution
will be destroyed by copious electron-positron pair production.
The second condition is violated by a classical electromagnetic
pulse whose energy $\mathcal{E}$ is small compared to the
characteristic frequency $\omega$. In quantum theory, such a pulse
will correspond to a state with a small mean number of photons.
Thus quantum fluctuations are large, and classical field theory
description is inadequate.

In the gravitational case, the first condition takes form of the
requirement that curvature should be $\ll 1$ (in Planck units). As
I have shown in \cite{bh1}, this condition is violated in the
collision spacetime of \cite{EG,YN} in a region relevant for the
horizon formation (see Section 2). The main purpose of this Letter
is to carry out the second check, by estimating the number of
gravitons participating in the collision (Section 3).

\section{Classical gravity picture and curvature estimates}
\label{class}

Let us focus on the ADD large extra dimension scenario
\cite{ADD} with fundamental
$D$-dimensional energy scale of gravity $\sim 1$ TeV. The
compactification radius $R$ is fixed so that at large distances we
recover the usual effective scale $M_\text{Pl}\sim 10^{19}$ GeV.
To avoid contradiction with the existing short-distance gravity
measurements \cite{short-distance}, we have to assume $D\ge 7$.
The Standard Model fields are localized on a 4-dimensional brane
embedded in the $D$-dimensional bulk.

If such or a similar scenario is realized in nature,
next-generation accelerators will be able to probe the
transplanckian regime of quantum gravity, colliding particles with
energy $E\gg 1$. (Here and below the $D$-dimensional Planck units
are used with $8\pi G=1$.) It is in such collisions that we may
hope to produce microscopic $D$-dimensional BHs which have mass
$M_\text{BH}\gg 1$ and are thus essentially classical. For
$M_\text{BH}\lesssim 1$ quantum gravity effects would be
significant without doubt.

The BH production process may always be considered as happening in
flat $D$-dimensional spacetime, because the Schwarzschild radius
of the created BH is much smaller than the compactification
radius:
\begin{equation}
R_S\sim E^\frac 1{D-3}\quad \ll\quad R\sim M_\text{Pl}^\frac
2{D-4}.
\end{equation}

In \cite{EG,YN} BH production was described by considering two
fast ($\gamma=E/m\gg 1$), point-like particles in a grazing
collision with an impact parameter $b$. Gravitational field of one
such particle in the limit $\gamma\to\infty$ has curvature
concentrated on the plane transverse to the direction of motion.
Introducing longitudinal coordinates $u=t-z$, $v=t+z$, and $D-2$
transverse coordinates $x_i$, the only nonzero components of the
Riemann tensor for the right-moving particle are \cite{EG}
\begin{equation}
\label{Riemann}
R_{uiuj}= \frac{E}{(D-4)\Omega}
\,\delta(u)\,\frac{\partial^2}{\partial x_i\partial
x_j}\left(\frac 1 {r^{D-4}}\right),
\end{equation}
($r=|x|$; $\Omega=\Omega_{D-3}$ is the volume of the unit
$(D-3)$-sphere.) This is the $D$-dimensional generalization of the
4-dimensional Aichelburg-Sexl shock wave spacetime
\cite{Pirani,AS,D'Eath,DH}.

This field should be superposed with the similar field of the
left-moving particle, shifted by $b$ in the transverse direction.
The resulting field is valid outside the region $u,v>0$, where the
colliding shocks start influencing each other. The metric in this
region should be found by solving Einstein's equation and remains
unknown even in the simplest $b=0$ case. Thus, BH formation may be
concluded only indirectly, using the CTS argument. The CTS's found
in \cite{EG,YN} are located in the known part of the spacetime.
They lie in the union of pre-collision parts of the shock planes
$u=0$ and $v=0$. In shape they look roughly like a union of two
throats narrowing around the particle worldlines in the far past
and glued together at an angle at the transverse collision plane
$u=v=0$ at radii $r\sim R_S$.

The validity of this classical Einstein gravity argument was
questioned by the present author in \cite{bh1} on the grounds that
curvature becomes large on the transverse collision plane at
$r\sim R_S$, i.e. in a region relevant for BH horizon formation.
Coordinate-invariant measure of curvature is provided by the
curvature invariant $(R_{\mu\nu\gl\gs})^2$. Notice that for a
single shock wave \rf{Riemann} this invariant vanishes, although
individual curvature components are large. This is not surprising,
since the Aichelburg-Sexl shock wave is a boost of a manifestly
low-curvature static Schwarzschild solution. However, when we add
a second particle, its left-moving shock wave will have large
$R_{vivj}$ components. As a result, nonzero contractions can be
formed, and we get
\begin{equation}
\label{Rsq} (R_{\mu\nu\gl\gs})^2 \sim E^{-\frac
2{D-3}}\,\gd(u)\,\gd(v)\qquad (r\sim R_S).
\end{equation}

To complete the curvature estimate, nonzero shockwave width
\begin{equation}\label{width}
    w \sim E^{-1}
\end{equation}
has to be taken into account. This width arises from the fact that
point particles are an idealization: in reality relativistic
quantum particles cannot be localized better then their wavelength
$\sim E^{-1}$. As a result, the delta functions in \rf{Rsq} have
to be smeared out on a scale $\sim w$. This gives the final
estimate \cite{bh1} of curvature at $r\sim R_S$:
\begin{equation}
\label{Rsq1}
(R_{\mu\nu\gl\gs})^2 \sim E^{-\frac 2{D-3}} w^{-2} \sim
E^{\frac{2(D-4)}{D-3}}\gg 1.
\end{equation}
This result means that higher curvature corrections to the
Einstein gravity may become important in the collision front and
significantly modify or even preclude BH formation \cite{bh1}.

\section{Graviton counting and quantum fluctuations in shock front}
\label{count}

The gravitational field of colliding particles was treated in the
previous section as classical. Was this justified?

To answer this question, we first of all have to decide about a
criterion when a quantum field can be considered classically. Such
a criterion is well known for the free electromagnetic field, and
in its strongest form it says that relevant photon occupation
numbers should be large \cite{BLP}. The weak form of the criterion
is to require that the {\it total} number of photons with relevant
frequencies be large. This condition guarantees that the
zero-point energy of quantum fluctuations is small compared to the
classical energy of the field.

I would like to apply an analogous criterion to gravitational
field. This should be possible for linearized gravity, when
deviation from the Minkowski metric is small:
\begin{equation}\label{linear}
    g_{\mu\nu}=\eta_{\mu\nu}+h_{\mu\nu},
    \qquad\left|h_{\mu\nu}\right|\ll 1.
\end{equation}
In this case $h_{\mu\nu}$ can be quantized as a free field, with
quanta being transverse gravitons\footnote{It should be noted
however that there are some indications that quantum gravity may
be very different from quantum field theory (e.g. because of the
``holographic principle" it may be required to satisfy, see
\cite{Hol} for a review). In such a case our analysis would not
apply.}.

Thus, I would like to count gravitons contained in the
gravitational field of colliding particles, say, of the
right-moving one. The presence of the other particle does not play
a role in this counting before the shock waves collide.

The standard shock wave metric \cite{EG}
\begin{equation}
\label{m1} ds^2=du\,dv-\frac {2 E}{(D-4)\Omega\,
r^{D-4}}\,\gd(u)\,du^2-dr^2-r^2\,d\Omega^2,
\end{equation}
corresponding to the Riemann tensor \rf{Riemann}, has blowing up
components and does not satisfy \rf{linear}. We will instead use
the metric \begin{equation} ds^2=du\,
dv-\left[1+\frac{(D-3)E}{\Omega\, r^{D-2}}\,
u\theta(u)\right]^2dr^2 -\left[1-\frac{E}{\Omega\, r^{D-2}}\,
u\theta(u)\right]^2r^2\,d\Omega^2 \label{m2},
\end{equation}
following from \rf{m1} by a coordinate transformation
\cite{D'Eath,EG,bh1}.
 Near the shock front, this metric can be approximated as
\begin{equation}\label{lin}
    ds^2\approx dx_\mu^2 -
    \frac{2{E}}{\Omega\,r^{D-2}}\,u\theta(u)\left[(D-3)dr^2-r^2\,d\Omega_{D-3}^2\right].
\end{equation}
We see that \rf{linear} is satisfied in the region of interest
$r\sim R_S$ and at $|u|\ll R_S$.

Since we want to count quanta, it is convenient to quantize the
graviton field $h_{\mu\nu}$ is one of the physical gauges. We will
use the transverse-traceless (TT) gauge, specified by conditions
\cite{MTW}
\begin{equation}\label{TT}
    h_{\mu0}=0,\quad h_{ik,k}=0,\quad h_{ii}=0.
\end{equation}
We can also take advantage of the fact that the shock profile is
slowly varying compared to the shock width \rf{width}. We can thus
work in plane wave approximation \cite{bh1}, neglecting transverse
derivatives of the metric:
\begin{equation}
    \label{nonzero}
    h_{ij}\approx h_{ij}(u,v).
\end{equation}

Our goal is to estimate the number of gravitons  contained in the
shock front at $r\sim R_S$, since this is the region relevant for
horizon formation. The $h_{ij}$ corresponding to \rf{lin} can be
written as
\begin{equation}
    \label{pwa}
    h_{ij} = \begin{pmatrix}
    -(D-3) C u\theta(u) && && &&\\
    && Cu\theta(u) && && \\
    && && \ddots && \\
    && && && Cu\theta(u) \end{pmatrix}.
\end{equation}
In the plane wave approximation the difference between polar and
Cartesian coordinates disappears. We can also neglect the
transverse dependence of $C$, so that it becomes a constant $\sim
R_S^{-1}$. After these simplifications, \rf{pwa} becomes precisely
of the form \rf{TT}, \rf{nonzero}.

Linearized gravity equations of motion satisfied by \rf{nonzero}
are just
\begin{equation}
    \label{eom}
    h_{ij,uv}=0.
\end{equation}
Quantization is performed by expanding into plane waves
\begin{equation}
    \label{quant}
    \hat{h}_{ij}=Z \int_{-\infty}^\infty \frac{dk}{(4\pi|k|)^{1/2}}
    \sum_{\alpha}\eps^{(\alpha)}_{ij}\,e^{ikz-i|k|t}\,\hat{a}_{k\alpha} + \text{h.
    c.},
\end{equation}
where the sum is over $D(D-3)/2$ independent symmetric traceless
graviton polarization tensors normalized by
$$
\eps^{(\alpha)}_{ij}\eps^{(\alpha')}_{ij}=\gd_{\alpha\alpha'}.
$$
The coefficient $Z$ has to be fixed so that the creation and
annihilation operators satisfy the standard commutation relations
\begin{equation}
    \label{comm}
    [\hat{a}_{k\alpha},\hat{a}^\dagger_{k'\alpha'}]=2\pi\, \gd_{\alpha\alpha'}\,\gd(k-k').
\end{equation}
 This field normalization can be performed starting from the linearized Einstein-Hilbert action in the TT
 gauge:
\begin{equation}
    \label{act}
    S=\frac 18\int d^{D}x\, h_{ij,\ga} h_{ij}{}^{,\ga}.
\end{equation}
In the plane wave approximation this becomes
\begin{equation}
    \label{actpw}
    S=\frac A8\int dt\,dz \left[(\partial_t h_{ij})^2-(\partial_z h_{ij})^2\right],
\end{equation}
where $A$ is the transverse area of the considered planar field
configuration, $A\sim R_S^{D-2}$ in our case. We see that $\half
A^{1/2} h_{ij}$ has the standard normalization of a 2-dimensional
massless scalar field. Thus we must take $Z=2A^{-1/2}$ for
consistency.

Now we are ready to count gravitons. First we expand $u\theta(u)$
in plane waves:
\begin{equation}
    \label{FT}
    u\theta(u)=-\frac 1{2\pi}\int dk\, e^{iku}\frac 1{(k-i0)^2}.
\end{equation}
The coherent state $|\Psi\rangle$ corresponding to a classical
solution $h_{ij}$ is characterized by the equation
\begin{equation}
    \label{Psi}
    \langle \Psi|\hat{h}_{ij}|\Psi\rangle = h_{ij}.
\end{equation}
From \rf{FT} and \rf{quant} we see that for our classical solution
\rf{pwa}
\begin{equation}
    \label{occ}
    \langle\Psi|\hat{a}_{k\alpha}|\Psi\rangle\
    \sim \frac C {Z k^{3/2}}
\end{equation}
(at $k>0$; the left-moving modes are of course in the vacuum
state). The mean occupation number of such quantum oscillator
state is
\begin{equation}
    \label{N}
    n_k=\langle\Psi|\hat{a}^\dagger_{k\alpha}\hat{a}_{k\alpha}|\Psi\rangle\sim
    \frac {C^2}{Z^2 k^3}
\end{equation}
Finally, the total number of quanta with energy $\sim \omega$ is
\begin{equation}
    \label{Ntot}
    N_{\omega}\sim \int_{k\sim \omega} dk\,n_k\sim
    R_S^{D-4}\omega^{-2}.
\end{equation}
This formula is only valid for $\omega\lesssim w^{-1}$ [see
\rf{width}] beyond which point the graviton spectrum sharply cuts
off.

Condition $N_\omega\gg 1$ gives the frequency range in which
graviton modes are classical:
\begin{equation}
    \label{classmod}
    \omega \ll \omega_{\max}\sim R_S^{(D-4)/2}.
\end{equation}
 For modes with $\go\gtrsim\go_{\max}$ occupation numbers
are small and quantum fluctuations become significant. This means
that classical gravity description is adequate only at distances
$|u|\gg\go_{\max}^{-1}$ from the shock front. It is easy to see
that this range is strictly smaller than the range $|u|\gtrsim w$
[see \rf{width}] necessary to resolve the full structure of the
classical gravitational field. In particular,
\begin{equation}\label{comp}
    N_{\go\sim w^{-1}}\sim E^{-\frac{D-2}{D-3}} \ll 1.
\end{equation}

\section{Discussion}

As we have seen, the gravitational fields involved in the problem
of BH production in particle collisions are of rather peculiar
nature. On the one hand, in the collision front curvature becomes
large, so that we expect transplanckian phenomena to take place
there (Section 2). On the other hand, precisely on the shock front
we find that gravitational field cannot even be considered
classical, in the sense that quantum fluctuations are large,
already for one shock (Section 3).

These elementary considerations strongly suggest that classical
Einstein gravity cannot be used to describe BH production
processes in TeV-scale gravity.  In particular, one cannot put too
much trust in the ``geometric cross section'' estimate for
microscopic black hole production \cite{BF,tHooft}:
\begin{equation}
    \label{geom}
    \sigma \sim \pi b_{\max}^2,
\end{equation}
where $b_{\max}$ is the maximal impact parameter for which a black
hole would be formed classically. The applicability of classical
gravity, against which I argued, is at the very foundation of this
estimate.

It is instructive to compare this dramatic situation with what
happens in fast collisions of macroscopic (say, solar mass) black
holes in $D=4$. In this case the shock wave width $w\sim r/\gamma$
is a purely classical finite-$\gamma$ effect \cite{D'Eath}.
Quantum wavelength contribution, which was dominant in the
microscopic case, is now absolutely negligible. The curvature in
the collision front at $r\sim R_S$ will be
\begin{equation}
\label{Rsqmacro} (R_{\mu\nu\gl\gs})^2 \sim E^{-4}\gamma^2 \ll 1
\end{equation}
for any reasonably imaginable value of $\gamma$. The number of
gravitons with characteristic frequency $\go\sim w^{-1}$ is
\begin{equation}
    N_{\go\sim w^{-1}}\sim E^2\gamma^{-2} \gg 1.
\end{equation}
These estimates should be contrasted with \rf{Rsq1} and \rf{comp}.
Not surprisingly, both conditions for applicability of classical
general relativity are satisfied by a huge margin, and the
problems discussed in this Letter do not arise here.

In conclusion, I would like to mention some of the recent
literature on the validity of the geometric cross section (see
\cite{Kanti} for a more complete list). That curvature may become
large in the collision front was previously suggested in
\cite{Hsu,Kan}, although without precise estimates. In \cite{Hsu}
it was stated without proof that the gravitational fields in
question contain many gravitons, which as we saw in Section
\ref{count} is not true at least in some part of the relevant
frequency range. The fact that the number of gravitons in the
shock front is small suggests that a perturbative treatment
similar to \cite{Kan} may be attempted in order to estimate the
resulting suppression of the geometric cross section. To avoid
possible confusion, it also has to be noted that my considerations
have nothing in common with Voloshin's exponential suppression
\cite{Vol} or arguments based on the ``generalized uncertainty
principle" \cite{Cav}.



\section*{Acknowledgements}
I would like to thank A.~Allahverdyan, J.~de Boer, L.~Cornalba,
 J.-W. van Holten, A.~Naqvi, A.~Polyakov, K.~Skenderis, and J.~Smit for useful discussions.
This work was supported by Stichting FOM.





\end{document}